\documentclass{kapproc} 

\usepackage{procps} 

\usepackage[dvips]{graphicx}

\upperandlowercase

\normallatexbib 

\begin{document}
\def\msun{${\rm M_{\odot}}$}

\articletitle{Gamma-ray Bursts and their Central Engines}


\author{Stephan Rosswog}
\affil{School of Engineering and Science\\
International University Bremen\\
Germany}
\email{s.rosswog@iu-bremen.de}

\begin{abstract}
Gamma-ray bursts are the most luminous and probably the most relativistic events in the universe.
The last few years have seen a tremendous increase in our knowledge of these events, but
the source of the bursts still remains elusive. I will summarise recent progress in this field 
with special emphasis on our understanding of the possible progenitor systems.
\end{abstract}


\section{Introduction}

Like some other spectacular discoveries such as the cosmic microwave background, gamma-ray bursts (GRBs) were discovered by accident. Meant 
to monitor the ``outer space treaty'', the American VELA satellites detected
in July 1967 an intense flash of gamma-rays of unknown origin. It took until
1973 before the first detected GRBs were published for the scientific community
(Klebesadel et al. 1973).\\
This caused furious research activity and a flurry of often very exotic theoretical models.
At the Texas conference in 1974, only one year after the first scientific publication,
Malvin Ruderman summarised the situation: ``The only 
feature that all but one (and perhaps all) of the very many
proposed models have in common is that they will not be the explanation of 
GRBs. Unfortunately, limitations of time prevent me from telling you which 
model is the exception....''. But his favourite model, accretion onto a black hole, is 
still a promising horse in today's race.

\section{Observations}


%
%



GRBs are short flashes of gamma-rays that outshine for short moment the whole rest of the 
gamma-ray sky. 
The BATSE instruments on board the Compton Gamma-Ray 
Observatory observed bursts at a rate of around one event per day (Paciesas et al. 1999). 
Opposite to initial expectations BATSE detected an inhomogeneous, but highly isotropic 
distribution of burst sources.
GRB lightcurves exhibit a tremendous variety ranging from single featureless spikes, 
over ``FREDS'' (fast rise exponential decay) to completely erratic sequences of pulses. The 
lightcurves vary on millisecond time scales, the shortest variability time scale so far 
being 0.22 ms for GRB920229 (Schaefer and Walker 1999).
The spectra of GRBs are non-thermal (see Figure 1) with high-energy tails extending up to
several GeV. The spectra can be accurately fitted with an exponentially smoothed, broken powerlaw, 
the so-called ``Band-function''
(Band et al. 1993). The ``knee'' where both powerlaws are joined is referred to as the break energy.
Typically break energies lie around several hundred keV. Generally, the lower
spectral energy index is compatible with a synchrotron origin of the radiation
(Cohen et al. 1997), there are however some bursts with a lower energy slope
steeper than predicted by the synchrotron model (Preece et al. 1998).
\begin{figure*}
\includegraphics[height=.4\textheight]{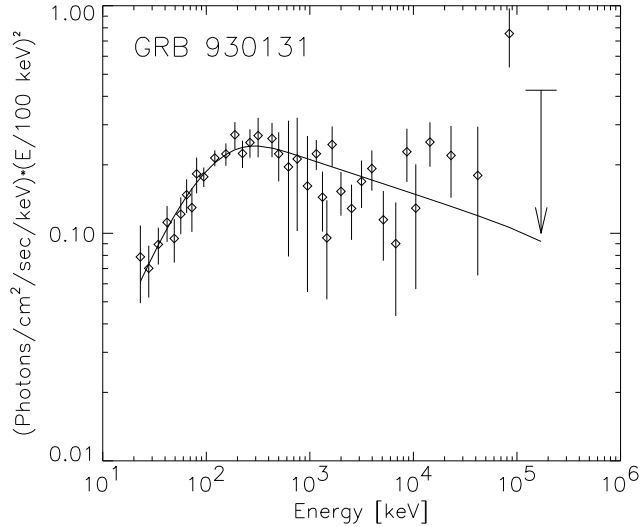}
{\caption{Spectrum of GRB930131 together with a synchrotron model fit (Bromm \& Schaefer 1999).}
\label{fig1}}
\end{figure*}
The duration distribution of GRBs is bimodal with a first peak at $\sim$ 0.2 s and a second one 
at $\sim$ 30 s. Bursts below (above) 2 s are referred to as ``short (long) bursts'' 
(Kouveliotou et al. 1993). 
Bursts of the short category exhibit predominately harder spectra 
than their long-duration cousins. The hardness is usually measured via the ``hardness ratio'',
the ratio of photon counts in a high and a low energy channel of BATSE.\\
The discovery of the first redshifts (Costa et al. 1997)  in GRB afterglows settled a long 
debate about whether they occur at cosmological distances or not. By now, the cosmological
origin of long GRBs is a well-established fact. Searches for the host galaxies 
found long bursts occurring predominantly in galaxies with active star formation (Bloom et al. 2002), 
pointing to a rather short-lived progenitor system, possibly related to the death of massive 
stars that die close to their birth-places. A GRB supernova connection had been 
suspected when in the localisation
error box of the BeppoSAX wide field camera of GRB980425 a supernova of type
Ib/c was found. By now, ``bumps'' have been detected in the optical afterglow 
lightcurves of several long GRBs (e.g. Bloom et al. 1999). These bumps were 
interpreted as the lightcurves from an underlying supernova that occurred roughly 
at the same time as the GRB. The idea of a long GRB supernova connection has
convincingly been corroborated by the detection of a very energetic supernova
that occurred temporally and spatially coincident with GRB030329 
(Hjorth et al. 2003, Stanek et al. 2003). Thus (at least some of the) long bursts seem to go 
along directly with a core-collapse supernova.\\
Short GRBs are usually assumed to be cosmological as well, but we are currently still 
lacking direct evidence in the form of afterglow observations. 
There are indications that the observed short GRBs occurred possibly 
at somewhat shorter distances (Mao et al. 1994). This would be consistent with
compact binary sources that merge relatively late in the age of the universe due to their long 
inspiral time (Fryer et al. 1999).\\
The fluences of typically $10^{-5}$ to $10^{-7}$ erg/cm$^2$ and cosmological redshifts observed 
so far imply isotropic burst energies of up to $\sim 4 \cdot 10^{54}$ ergs for GRB990123, 
corresponding to twice rest mass energy of the sun in gamma-rays alone. This
``energy crisis'' pointed to beamed emission. By now, in several GRBs achromatic breaks
have been identified from which (under certain assumptions) jet opening angles
of $\sim 5^o$ have been derived (Frail et al. 2001). Taking beaming angle corrections into account,
the energy requirements are alleviated, the initially huge range of isotropic energies ranging from 
a few times $10^{51}$ to $4 \cdot 10^{54}$ ergs collapses to a narrow distribution around
$\sim 5 \cdot 10^{50}$ erg. If true, this allows the use of GRBs as standard candles
to infer, for example, star formation in the  high redshift universe
(e.g. Lloyd-Ronning et al. 2002). 
\\
We briefly want to mention two exciting, but currently still controversial
topics: X-ray emission lines and polarisation of the prompt GRB-emission.
X-ray emission features have been reported for several X-ray afterglows
(e.g. Piro et al. 2000). The confidence level for these detections, however, is rather
low. The energy contained in these lines is very large, $\sim 10^{49}$
erg. Since the efficiency to produce an X-ray emission line cannot exceed 1
$\%$ the burst would have to have $\sim 10^{51}$ ergs in X-rays (Ghisellini et
al. 2002), which sets a strict lower limit on the burst energy that is hard to reconcile
with the results of Frail et al. (2001).\\
Using RHESSI results Coburn and Boggs (2003) reported on a high linear polarisation of the
prompt emission of GRB021206, $\Pi=0.8 \pm 0.2$. This could point to the interesting
possibility of magnetic fields being advected from the central source rather
than being generated in shocks. A re-analysis of the same data, however,
found a no clear indication of for polarisation of GRB021206 (Rutledge and Fox, 2003).

\section{The fireball model}

The above listed observational facts impose constraints on the 
progenitor system(s) responsible for gamma-ray bursts.\\
The first one, often referred to as the ``compactness-problem'' comes from the non-thermal GRB 
spectra and was originally used as an argument against a possible cosmological origin 
of GRBs (Ruderman 1975, Schmidt 1978). The millisecond variability of the lightcurves points
to a compact source for the GRB, as from simple causality arguments the dimension of the
system should be $D < c \cdot \delta t \approx 300$ km. But if an energy of $\sim 10^{51}$ erg 
in $\sim$ 1 MeV photons is released in such a small volume the optical depth to pair
creation must be huge: $\tau \sim 10^{13}$ and therefore one would expect a thermal spectrum,
in blatant contrast to the observations. The way out of this dilemma is relativistic motion
(Paczynski 1986, Goodman 1986). If the emission is coming from material moving with Lorentz factor
$\Gamma$ towards the observer, the source can be larger by a factor of $\Gamma^2$. Moreover the
photons in the local frame are softer by a factor of $\Gamma$. A detailed analysis (Lithwick and 
Sari 2001) shows that the optical depth can be reduced by effects of relativistic motion by a factor 
$\sim \Gamma^{6.5}$ where the exact value of the exponent depends on the GRB spectrum. The analysis 
of Lithwick and Sari yields lower limits on GRB Lorentz factors of several hundreds, the highest 
Lorentz factors in the universe.\\
A second hard constraint is the so-called ``baryon loading problem''. The attainable bulk Lorentz factor
is determined by the ratio of available energy and rest mass energy. If an energy of $10^{51}$ erg is 
available the fireball cannot contain more than $10^{-5}$ \msun in baryonic material, otherwise the 
required Lorentz-factors will not be reached. This poses a hard problem for central engine models: how can 
a stellar mass object pump so much energy into a region that is  essentially devoid of baryons?

The above reasoning has led to the ``fireball'' internal-external shocks model.
This model is rather independent of the nature of the central engine. The latter one is just required to 
produce highly relativistic outflow, either in the form of kinetic energy or as Poynting flux.
The radiation is produced in (collisionless) shocks. These can either occur due to interaction of the outflow
with the circumstellar material (``external shocks'') or due to interactions of different portions of the outflow
with different Lorentz-factors, so-called ``internal shocks''. According to the fireball model the GRB
is produced via synchrotron radiation (at least in the most simple model version) in internal shocks.
The efficiency of internal shocks depends on the variability within the outflow (Kobayashi et al. 1997). 
A lot of energy remains as kinetic energy in the ejecta and can later be dissipated via external shocks to 
produce the much longer lasting multi-wavelength afterglow. Such an afterglow had been predicted before its
observation by several researchers (Paczynski and Rhoads (1993), Meszarosz and Rees (1993, 1997), 
Vietri (1997)). The simplest afterglow model with an adiabatic, relativistic blast wave (Blandford and McKee 1976) 
and synchrotron radiation, seems to fit the bulk of afterglow observations reasonably well (for a detailed discussion see Piran (1999) and Meszaros (2002)).\\
Detailed investigations of Kobayashi et al. (1997) showed that the GRB lightcurve reflects essentially
the activity of the inner engine that produces the relativistic outflow. That means that the engine
itself has to produce an erratic sequence of pulses and, whatever the progenitor is, it should be able 
to produce, depending on its specific system parameters a large variety of different outcomes. This
result seems to rule out ``one-bang'' models, e.g. simple phase transitions in neutron stars.

For an alternative to the fireball model see Dar and De Rujula (2003).

\section{Models for the Central Engine}

Many many GRB models have been suggested over the years and it is impossible to do justice to all of them here.
I will therefore just pick out a few that I consider to be particularly interesting.\\
When discussing the central engine it is worth keeping in mind that the overall burst duration, $\tau$, is 
-both for long and for short bursts- {\em very long} in comparison to the variability time scale, $\delta t$, 
on which the energy output changes substantially, $\delta t/\tau \ll 1$. This requires the engine to provide (at
least) these two different time scales. One (but not the only) reasonable possibility is that the central engine
consists of a new-born, stellar mass black hole surrounded by an accretion disk. In this view the dynamical
time scale close to the hole sets the variability while the viscous accretion time scale sets the overall burst 
duration. The hole/disk masses then distinguishes between long and short duration bursts. While this is 
certainly reasonable, I want to stress here that plausible models do not necessarily have to involve a black hole,
examples of such alternative models are Usov's (1992) highly magnetised pulsar formed in an accretion induced 
collapse or the temporarily stabilised central object resulting in a neutron star coalescence, see below.\\
One important question is whether there is a single progenitor that produces (depending on say its initial
conditions like the rotation rate) either a short or a long GRB or whether the two burst classes result form
two different progenitor systems. An example of a model that produces both long and short GRBs from one progenitor
has been suggested by Yo and Blackman (1998). In their model the GRB is driven by dissipation of Pointing 
flux extracted from a young, highly magnetised millisecond pulsar formed in an accretion-induced collapse of 
a white dwarf. If the initial rotation is above a critical rotation rate, the spin-down time scale is 
determined by gravitational wave emission leading to sub-second spin-down, otherwise its is governed by 
electromagnetic dipole emission and leads to a much longer duration ($\gg$ 1 s).\\
There are, however, reasons to assume that the two GRB classes are caused by different progenitor systems since
there are, apart from the duration further substantial differences. First, as mentioned above, short GRBs seem to be 
systematically harder than their long-duration cousins.  Moreover their lightcurves exhibit fewer sub-pulses 
and they show a different temporal spectral evolution (Norris et al. 2000). 
Apart from that the spectral break energies
of short bursts seem to be larger than for long bursts (Paciesas et al. 2000). As the spectral break energy is 
sensitive to the cosmological redshift and the Lorentz-factor of the ejecta producing the burst, the higher 
spectral break energies could mean that either short GRBs have higher Lorentz-factors or/and that their population 
is closer, i.e. they have, on average, a lower redshift. Both of these possibilities are compatible with compact 
binary mergers: due to their inspiral time they could occur relatively late in the age of the universe and therefore
be at lower redshifts than long GRBs. Moreover, they are not engulfed in a stellar mantle like in the case of a stellar
collapse and the high rotation velocities that naturally occur will centrifugally evacuate the region around the 
binary rotation axis. This fact will make it easier to obtain higher Lorentz-factors without being slowed down
by a stellar mantle that has to be penetrated first. Moreover, the two classes seem to be differently distributed 
in space (Cohen, Kolatt and Piran 1994), consistent with short GRBs being closer than their long-duration cousins.
All this hints to two different progenitor systems.

\subsection{Long-soft Bursts}

Around 70 $\%$ of the observed GRBs are the long-soft type. They typically occur at cosmological redshifts of $z \sim 1$
and are believed to be beamed with half-opening angles of $\sim 5 ^o$. Thus, we see on average one burst per 300 beamed 
GRBs with random jet orientations, i.e. the true rate of long GRBs is around 1000 per day in the universe.

\subsubsection{Collapsars}
\begin{figure*}
\includegraphics[height=.38\textheight]{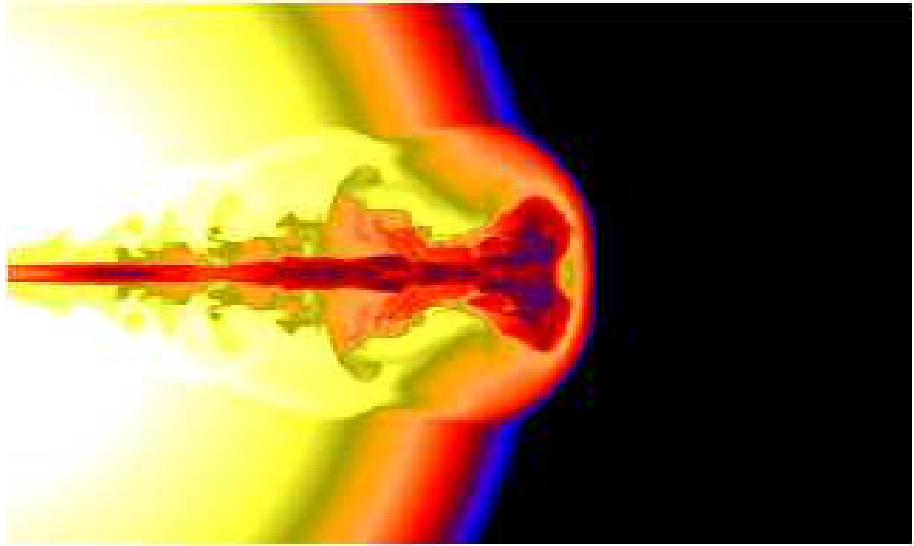}
{\caption{Density distribution of a collapsar jet breaking through the stellar surface. 
From Zhang, Woosley and Heger 2003.}
\label{fig1}}
\end{figure*}
The probably most popular GRB-model is the ``Collapsar''/``failed supernova'' model (Woosley 1993, MacFadyen and Woosley 1999), 
sometimes also referred to as ``hypernova'' (Paczynski 1998).\\ 
The exact supernova mechanism is despite intense research not known with certainty, but the general believe is 
that the huge neutrino luminosities from the newborn proto-neutron star will revive the shock that got stalled 
in the stellar mantle and will drive the explosion. If, however, a black hole rather than a neutron star is formed 
early on, the neutrino source is essentially shut off and the supernova might ``fail''. For a non-/slowly rotating
star the stellar matter can essentially unhindered disappear in the event horizon of the new-born black hole.
If the angular momentum of the star is too high, a disk will form at too large distances to be hot enough for efficient
neutrino emission. For intermediate values of angular momentum, however, a more compact, hot accretion disk may form
close to the hole. In this case an accretion- rather than neutrino-driven supernova explosion might occur.
Although there is no generally accepted mechanism for its explanation there
seems to be a common connection between accretion disks and jets. Like in many
other astrophysical environments such as active galactic nuclei and young
stellar objects the formation of a jet is also plausible in the collapsar case. It might be powered by energy pumped into
the polar region above the hole either via the annihilation of neutrinos emitted from the inner, hot parts of the 
accretion disk or via MHD mechanisms. Initially the density along the original rotation axis is too high for any
plausible jet to overcome the ram pressure of the infalling material. But after a few seconds a  channel along the 
rotation axis opens up allowing later on for jet propagation through the star. This channel is expected to 
collimate the escaping jet. If the hole, however, has received too large a kick at birth the energy deposition 
will not be along the evacuated channel and the jet may be spoilt.\\
Apart from having the right amount of angular momentum the progenitor star is required to have lost its envelope, 
since a jet powered for the typical duration of a long burst, $\sim 20$ s, could not penetrate a giant star, it would 
rather dissipate its energy in the stellar envelope. To penetrate say a red 
supergiant it would take around 1000 s and it is hard to see why the burst after having been on for so long should then 
suddenly shut down within 20 s after breaking through the stellar surface. Therefore, the progenitor stars is expected 
to have lost its envelope and the GRB is expected to go along with a supernova of type I (i.e. core-collapse but no 
hydrogen lines). Resulting from very massive stars,
collapsars would occur naturally in star forming regions and, as observed, to be associated with a core-collapse supernova
of type Ib/c.

The first simulations of the collapsar scenario have been performed using 2D Newtonian, hydrodynamics 
(MacFadyen $\&$ Woosley 1999) exploring the collapse of helium cores of more than 10 \msun. 
 In their 2D simulation MacFadyen $\&$ Woosley found the jet to be collimated by the 
stellar material into opening angles of a few degrees and to transverse the star within $\sim 10$ s. The 
accretion process was estimated to occur for a few tens of seconds. In such a model variability in the 
lightcurve could result for example from (magneto-) hydrodynamic instabilities in the accretion disk that 
would translate into a modulation of the neutrino emission/annihilation processes or via Kelvin-Helmholtz 
instabilities at the interface between the jet and the stellar mantle.\\
These initial, Newtonian simulations were plagued by highly superluminal motions. Aloy et al. (2000) have
improved on this problem using a special relativistic hydrodynamics code to follow
jet propagation through a progenitor star. Recently Zhang et al. (2003) have simulated the propagation 
and breakout of a jet through a Wolf-Rayet star using a 3D, special relativistic grid code. They particularly
addressed the question of jet stability going from 2D to 3D and found the gross jet features to be robust
against this change. Moreover, they found that even a slowly precessing jet (say a degree per second) is
able to penetrate the star and emerge relativistically at its surface. If the jet precesses faster, it 
is dissipated inside the stellar envelope. At outbreak the jet is surrounded
by a mildly relativistic cocoon that will give rise to a transient signal. If the outbreak is seen at a larger
angle with the jet axis it may appear as a so-called ``X-ray flash'' (Heise 2003)

Despite its obvious success there remain several open questions. First and most important is the progenitor
question. Do stars with the right amount of angular momentum at collapse exist at all? If state of the art
progenitor models in which angular momentum loss through magnetic torques and wind losses are 
accounted for (Heger and Woosley 2003) are taken at face value, then the collapse will not result in a 
sufficiently massive disk to produce a collapsar. However, as the progenitor uncertainties are substantial that does not
necessarily mean a killer argument for collapsars. Another essentially unsolved problem is the question
how the jet is launched. In current numerical models the jet {\em formation} is not modelled, jets are put in
by hand and then their propagation through the star is followed. Further uncertainties 
come from the numerical resolution. As short-wavelength perturbations grow fastest in a Kelvin-Helmholtz instability
the question remains whether current models can already resolve the smallest physically important scales. But
this will certainly be addressed in future work. As in many current astrophysical simulations
the first collapsar models ignored the influence of magnetic fields. Currently a lot of effort is put into 
MHD-simulations of this event (see e.g. Proga et al. 2003).

\subsubsection{Supranova}

Vietri and Stella (1998) have suggested a two stage process to create a GRB (``supranova''). 
 The massive star collapses in a ``traditional'' supernova to form a rotationally 
supported neutron star that is beyond its non-rotating upper mass limit. They assume the star to spin down via magnetic dipole
radiation. Using a typical neutron star magnetic field of $10^{12}$ G they estimate a time scale of several
years before the  ``supra-massive'' configuration collapses into a black hole. The supernova is expected to leave a 
$\sim 0.1$ \msun  accretion disk in orbit around the neutron star and, once the neutron star collapses to form a black 
hole, the disk might help to extract the rotational energy of the new-born black hole and to launch the GRB.\\
The strengths of the model are its natural connection to supernovae and star formation and
that the supernova remnant would have enough time to form iron via the decay of
nickel and cobalt to possibly produce the claimed iron lines. Moreover, it is expected to be a baryon-clean environment. The model is, however, very sensitive
to the fine tuning of parameters. Moreover, GRB030329 places a rather strict limit of a few hours on the delay between the SN and the GRB
and thus rules out the supranova model for at least this particular burst.

\subsubsection{Fragmenting Core-collapse}

Another two stage process, the fragmentation of a core collapse supernova has been suggested recently by Davies et al.
(2002). The main idea is that, analogous to the fragmentation of a collapsing molecular cloud into several stars,
a sufficiently rapidly rotating progenitor star might undergo fragmentation and form several ``lumps'' rather than a
clean and more or less spherically symmetric neutron star. The various fragments would then be driven towards 
coalescence via gravitational wave emission and would emit a characteristic chirp signal during inspiral.
Once coalesced they will form a ``standard GRB central engine'', a new-born black hole and an accretion disk. 
Generally the black hole will receive a kick at birth, a successful GRB-jet can only be launched if the received kick velocity
is small. For plausible system parameters a delay of several hours between supernova and GRB can be obtained. This is well within the
current limits set by GRB030329. This model shares with the two previously discussed models its connection with star formation
regions and core-collapse supernovae. The exact conditions under which such
a fragmentation can occur and the fraction of core-collapse systems that will
undergo fragmentation remains to be explored in future work.

\subsection{Short-hard Bursts}

As outlined above, the class of short-hard GRBs probably results from a different central engine than long bursts.
Compact binaries, either two neutron stars \cite{eichler89} or a neutron star and a low mass black hole \cite{paczynski92}, are the most promising 
candidates for the progenitor of this class. Their gravitational binding energy of a few times 
$10^{53}$ erg is expected to easily satisfy the energy requirements of a short GRB (the exact energy in the burst 
is at present not known since so far no redshift has been detected). As the neutron star dynamical time scale is 
$\tau_{dyn}= (G \bar{\rho})^{-\frac{1}{2}} \sim 3\cdot 10^{-4}$ s there should be no problem in providing even the
shortest time scales observed in bursts. Moreover, the merger remnants are attractive 
GRB central engines as the black hole (either preexisting or forming after a neutron star binary coalescence)
is guaranteed to be rapidly rotating, either from the angular momentum that has been fed into the hole 
in the form of neutron star debris  or from the orbital motion of a neutron star binary. A soft equation of state 
leads to more compact neutron stars, therefore a neutron star binary system will have higher orbital velocities 
at contact and thus a more rapidly rotating black hole will form. The energy stored in  the black hole rotation can then possibly 
be extracted, e.g. via the Blandford-Znajek mechanism (1977).\\
With the recent discovery of another highly relativistic binary pulsar (Burgay et al. 2003) the estimates 
for neutron star merger rates have, once again, been increased. With the most likely rates estimated to 
be as high as $\sim 2 \cdot 10^{-4}$ per year and galaxy (Kalogera et al. 2003), neutron star binary 
mergers will have no problem to account for all short GRBs, even if substantial beaming is involved.\\
Neutron star black hole mergers are most often considered to be just a variant of the binary neutron
star theme in the sense that they will also produce a black hole plus torus system. It is worth pointing
out, however, that the dynamics of this merger process is much more complex than in the neutron star merger
case as it is governed by the interplay of gravitational radiation (trying to {\em decrease} the orbital 
separation), mass transfer (trying to {\em increase} the orbital separation) and the reaction of the stellar
radius to mass loss. The dynamics of this process is very sensitive to the mass ratio of the 
binary components which is much less constrained than in the binary neutron star case.

\subsubsection{Neutron Star Binary Mergers}

Neutron star binaries have for a long time been {\em the} model for the central 
engine of GRBs (Goodman et al. 1987, Eichler et al. 1989, Narayan et al. 1992). 
With the connection of long GRBs with supernovae being well established by now,  
compact binary mergers are these days  not usually considered to be responsible
for long duration GRBs, but they are still the leading model for short GRBs.\\
The coalescence is an intrinsically three-dimensional phenomenon and therefore 
analytical guidance is rare (although very welcome) and one has to resort to large 
scale numerical computations. Additional complications arise from the fact that there is 
almost no field of astrophysics that does not enter at some stage during the 
coalescence process: the last stages and the merger are certainly dominated by 
strong-field general relativistic gravity, the neutron star material follows 
the laws of hydrodynamics, particle physics enters via possible condensates of 
``exotic'' matter in the high-density interiors of the neutron stars and the 
copiously produced neutrinos in the hot and dense neutron star debris, questions 
concerning element formation require detailed information on nuclear structure and
reactions (often far from stability) to be included and also magnetic
fields might play a decisive role since they may, via transport of angular
momentum, determine whether and/or when the central, coalesced object collapses
into a black hole. Moreover, they are expected to be amplified in the complex fluid 
motions following the merger to field strengths where they become dynamically important.\\
Due to this complexity current investigations follow one of two
``orthogonal'' lines: either ignoring microphysics, resorting to the simplest
 equations of state (EOS), polytropes, and thereby focusing on
solving the complicated set of general relativistic fluid dynamics equations
(or some approximation to it; e.g. Shibata and Uryu 2000, Oechslin et al. 2001) or,
along the other line, using accurately treatably (Newtonian) self-gravity of the 
fluid and investigating the influences of detailed microphysics and relating the 
merger event to astrophysical phenomena (e.g. Ruffert et al. 2001, Rosswog et al. 2002).

\begin{figure*}
\hspace*{-2cm}\includegraphics[height=.25\textheight]{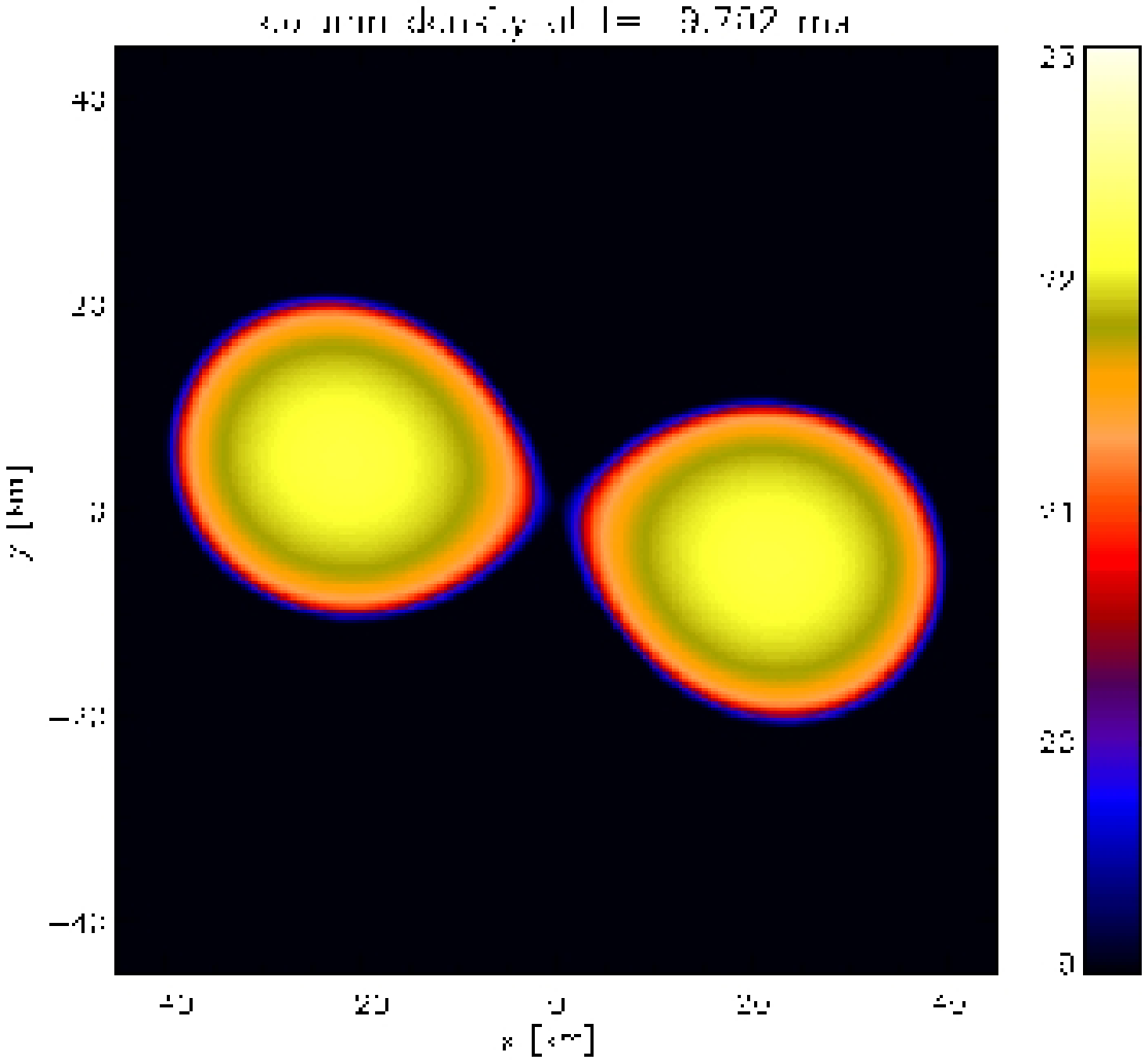}\includegraphics[height=.25\textheight]{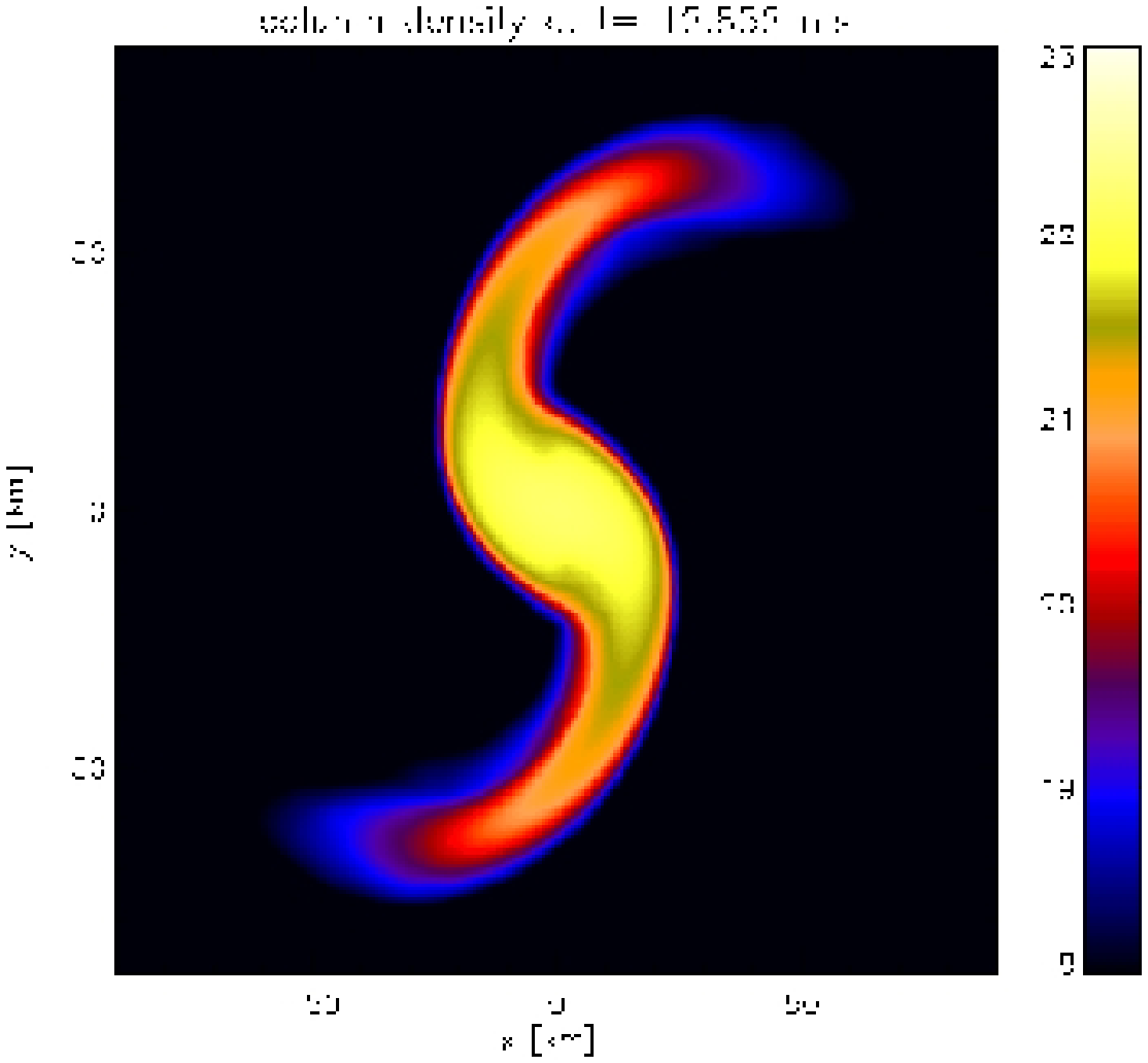}\includegraphics[height=.25\textheight]{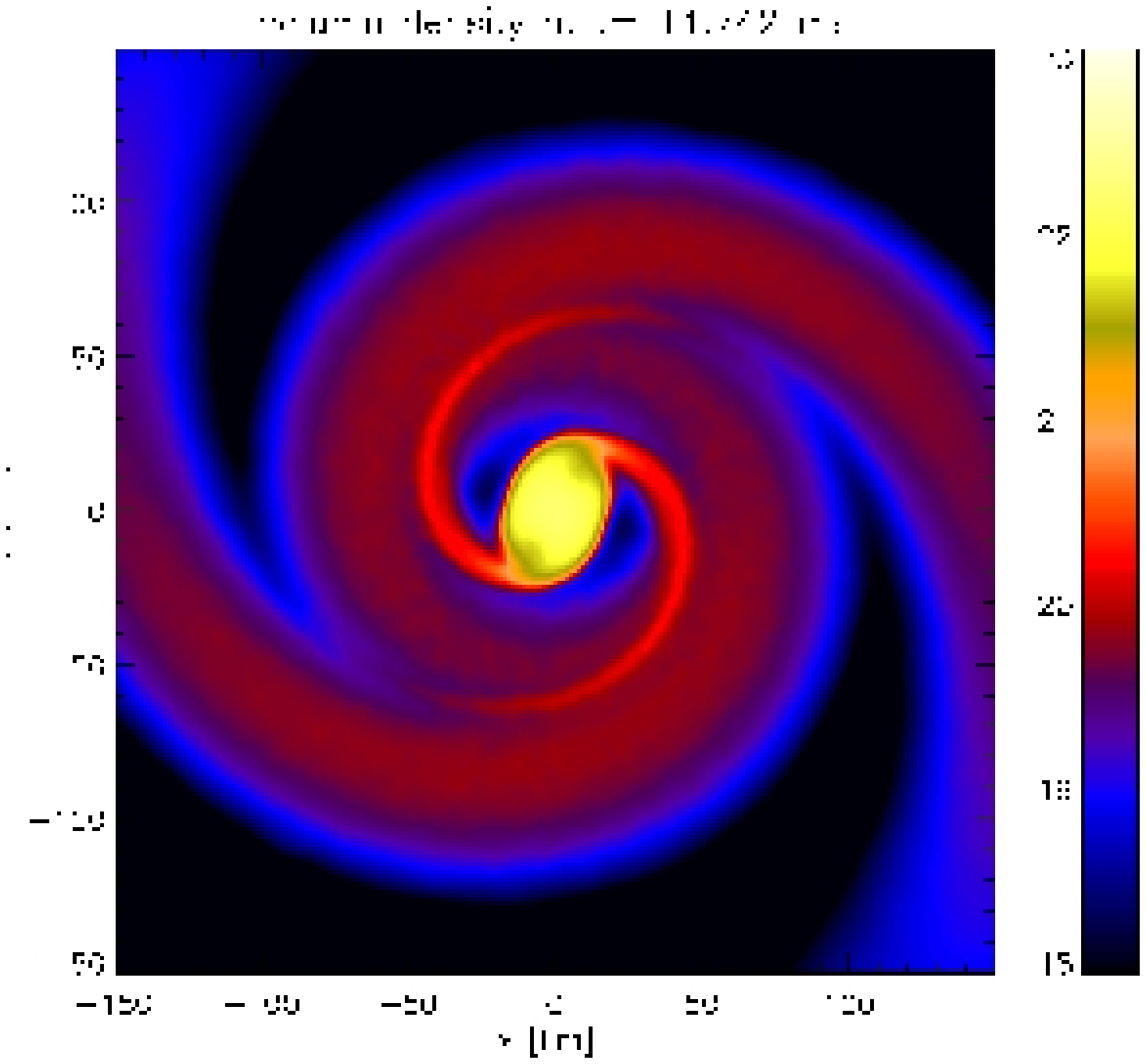}
{\caption{Coalescence of a corotating neutron star binary system (1.4 \msun each star). 
Colour-coded is the column density, the axes are in kilometers. The simulations are described 
in detail in Rosswog et al. 2002.}
\label{fig1}}
\end{figure*}

How can the available energy now be transformed into outflowing
relativistic plasma after such a coalescence event? 
A fraction of the energy released as neutrinos
is expected to be converted, via collisions,
into electron-positron pairs and photons and, if converted in a region of
low baryon density, this will give rise to a relativistic wind. Alternatively,
strong magnetic fields anchored in the dense matter could convert the
available energy of the system into a Poynting-dominated outflow.\\

\centerline{Neutrino annihilation}

The merger remnant emits neutrinos in copious amounts. The total neutrino 
luminosities are typically around $2 \cdot 10^{53}$ ergs/s with electron-type
anti-neutrinos carrying away the bulk of the energy. Typical neutrino energies
are not too different from those of type II supernovae, electron neutrinos 
are around 8 MeV, their anti-particles around 15 and the heavy lepton 
(anti-)neutrinos have energies around 22 MeV (Rosswog and Liebendoerfer 2003).\\
Neutrino annihilation is an attractive process to launch a fireball as
neutrinos can leave the hot remnant easily and can deposit their energy
possibly in a region devoid of baryons to avoid prohibitive baryon loading.
The centrifugally evacuated funnel region above the remnant is an
ideal place for this deposition to occur because it is close to
the central energy source (the energy deposition rate scales roughly
with the inverse fourth power of the distance) but contains only a
small number of baryons (e.g. Ruffert et al. 1997,
Rosswog and Davies 2002). Here  the $\nu \bar{\nu}$ energy
flux can be transformed into a radiation-dominated fluid with a high entropy per
baryon. The thick disk geometry of the remnant with its steep density
gradients in the radial direction does not allow for lateral
expansion.  The only escape route is along the initial binary rotation
axis. The disk geometry is therefore responsible for channelling the
relativistic outflow into a pair of anti-parallel jets. This mechanism
is similar to that envisaged by MacFadyen \& Woosley (1999) for the
collapsar scenario, but offers the advantage that the jets in the merger case
do not have to pierce through a surrounding stellar mantle.\\
\begin{figure*}
\includegraphics[height=.48\textheight,angle=90]{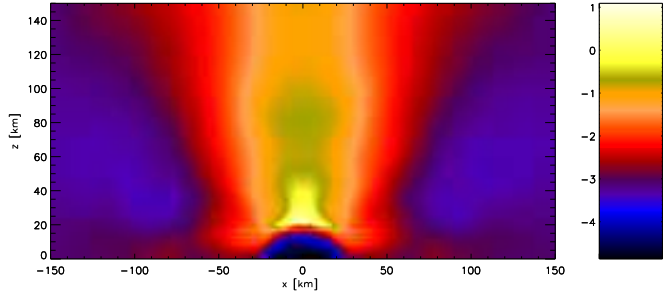}
{\caption{The annihilation of neutrino-antineutrino pairs above the remnant of a neutron star merger drives
relativistic jets along the original binary rotation axis (only upper half-plane is shown). 
The x-axis lies in the original binary orbital plane,
the dark oval around the origin is the newly formed, probably unstable,
supermassive neutron star formed in the coalescence. Colour-coded is the
asymptotic Lorentz-factor. Details can be found in Rosswog et al. 2003.}
\label{jet}}
\end{figure*}
Neutrino annihilation from full merger simulations has been calculated 
by Ruffert et al. (1997) and Rosswog et al. (2003).
Popham et al.(1999) have studied the annihilation for the case of steady state accretion disks.
The resulting bulk Lorentz-factor is determined by the annihilation energy deposited per rest mass energy.
This quantity is displayed in Figure \ref{jet} (for details see Rosswog et al. 2003).
The encountered Lorentz-factors range from around 10 (this is a lower limit as 
finite numerical resolution leads us to overestimate the funnel density and therefore to 
underestimate the Lorentz-factor) to a few times $10^4$
for extreme cases (see Rosswog et al. 2003 for details), so Lorentz 
factors of several hundred should not pose any problem for this mechanism.\\
The total energy in the relativistic outflow, however, is only of order 
$10^{48}$ ergs and therefore moderate by GRB standards. Ruffert et al. (1997)
find somewhat higher values, but agree that neutrino annihilation can only
power a relatively weak burst. To account for the observed luminosities of
a short GRB, the relativistic outflow has to remain well-collimated.
One such collimation mechanism (Levinson and Eichler 2000) relies on the beaming of the relatively 
weak, relativistic jet by the ram-pressure of the non-relativistic, but powerful
baryonic wind that is blown off the merger remnant via 
neutrinos (this is very similar to the neutrino-driven wind of a new-born 
proto-neutron star). Using a theoretical neutron star mass distribution
Rosswog and Ramirez-Ruiz (2003) have calculated the resulting beaming angle
and the apparent luminosity distribution. They find a broad distribution of apparent 
luminosities (typical values for isotropized luminosities of $\sim 5 \cdot 10^{50}$ erg/s)
with average jet opening angles of $\sim 5^o$. These results are compatible
with both the fluences observed for short GRBs and with the estimated neutron star merger rates.

\bigskip
\bigskip

\centerline{Magnetic mechanisms}

\begin{figure}
  \includegraphics[height=.38\textheight]{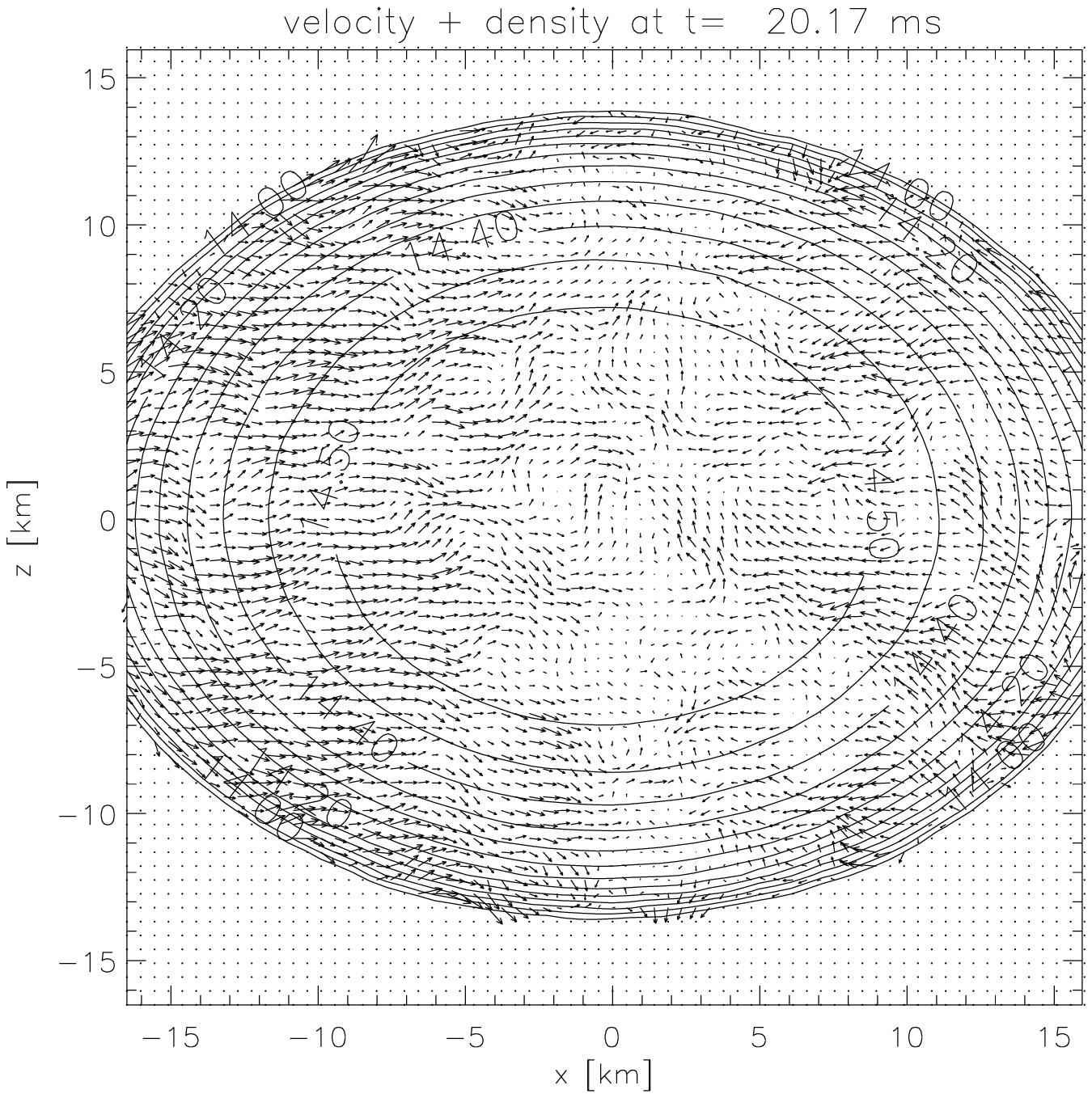}
  \caption{Velocity field (space-fixed frame) inside the central object of
the remnant of a neutron star coalescence. The labels at the contour lines 
refer to log($\rho$), typical fluid velocities are $\sim 10^8$ cm/s.}
\label{vel}
\end{figure}
Neutron stars are endowed with strong magnetic fields, typically of order
$10^{12}$ G. Even the sun with its comparatively moderate magnetic fields
and fluid motions produces a large spectrum of different magnetic activities
with sometimes violent outbreaks. Thus it is natural to expect scaled
up versions such activity in an event as vigorous as a neutron star merger.
Several magnetic GRB scenarios have been discussed over the years, e.g.
by Usov (1992, 1994), Narayan et al. (1992), Duncan and Thompson (1992), 
Thompson and Duncan (1994), Meszaroz and Rees (1997), Katz (1997), 
Kluzniak and Ruderman (1998) and Rosswog et al. (2003).\\
I want to mention here three different possibilities: first, that the central
object formed in the coalescence remains stable for (at least) a short time of the order
a second before collapsing to black hole. During this time the seed magnetic fields 
can be amplified drastically and the central object can, as some kind of a ``superpulsar'',
launch a short lived relativistic wind. Second, similar field amplification processes are expected to occur
in the accretion torus around the central object. Finally, if the central object 
should collapse immediately into a black hole (i.e. before its rotational energy 
can be extracted by the superpulsar mechanism) its rotational energy can be extracted
via a magnetic coupling to the accretion torus. The latter two processes have been discussed 
in the literature (e.g. Blandford and Znajek 1977, Narayan et al. 1992, Rosswog et al. 2003).
We will therefore draw attention to the first possibility.\\ 
The central object of the remnant is rapidly differentially rotating 
(see Rosswog and Davies 2002 for rotation curves) with rotational periods ranging
from $\sim 0.4$ to $\sim 2$ ms. Differential rotation is known to be very efficient 
in stabilising stars that are substantially more massive than their non-rotating 
maximum mass. For example, Ostriker and Bodenheimer (1968) constructed differentially
rotating white dwarfs of 4.1 \msun. A recent investigation analysing 
differentially rotating polytropic neutron stars (Lyford et al. 2002) finds 
it possible to stabilise systems even beyond twice the typical neutron star
mass of $ 2.8$ \msun. The exact time scale of this stabilisation is 
difficult to determine, as all the poorly known high-density nuclear physics
could influence the results, but for this process to work, only a time scale
of about a second is needed (some authors have estimated time scales of up to years
\cite{shibata99}).\\
It is worth pointing out that the fluid flow in our calculations of the merger 
(Rosswog et al. 2003) never becomes axisymmetric during the simulation and 
therefore Cowlings anti-dynamo theorem does not apply here. When the 
surfaces of the neutron stars come into contact, a vortex sheet forms between them 
across which the tangential velocities exhibit a discontinuity. This vortex sheet is
Kelvin-Helmholtz-unstable. These fluid
instabilities lead complex flow patterns inside the central object of
the merger remnant. In the orbital plane they manifest themselves as strings
of vortex rolls that may merge (see Fig. 8 in Rosswog and Davies 2002). An example 
of the flow pattern perpendicular to the orbital plane is shown in  Fig. \ref{vel}.
This pattern caused by fluid instabilities exhibits ``cells'' of size $l_c \sim$ 1 km
and velocities of $v_c \sim 10^8$ cm/s. Moreover, we expect neutrino emission to drive
convection as in the case of proto-neutron stars. The neutrino optical depth within
the remnant drops very steeply from $\sim 10^4$ at the centre to the edge of the central object (see Fig. 11 
in Rosswog and Liebendoerfer 2003). For this reason the outer layers loose neutrino energy, entropy and 
lepton number at a much higher rate than the interior. This leads to a
gradual build-up of a negative entropy and lepton number gradient which will
drive vigorous convection (e.g. Epstein 1979). We expect this to set in
after a substantial fraction of the neutrino cooling time (i.e. on time scales longer than
our simulated time) when a lot of the thermal energy of the remnant has been radiated away.
With neutrinos as dominant viscosity source we find viscous damping time scales of several
tens of seconds, i.e. much longer than the processes we are interested in (see below).\\
We expect an efficient $\alpha-\Omega$-dynamo to be at work in the merger remnant.
The differential rotation will wind up initial poloidal into a strong toroidal field
(``$\Omega$-effect''), the fluid instabilities/convection will transform toroidal
fields into poloidal ones and vice versa (``$\alpha-$effect''). Usually, the
Rossby number, $Ro \equiv \frac{\tau_{\rm rot}}{\tau_{\rm conv}}$ is adopted as a measure
of the efficiency of dynamo action in a star. In the central object we find Rossby
numbers well below unity, $\sim 0.4$, and therefore expect an efficient amplification of
initial seed magnetic fields. A convective dynamo amplifies initial fields
exponentially with an e-folding time given approximately by the convective 
overturn time, $\tau_c \approx 3 $ ms; the saturation field strength
is thereby independent of the initial seed field (Nordlund et al. 1992).\\ 
Adopting the kinematic dynamo approximation we find that, if we start with a 
typical neutron star magnetic field, $B_0= 10^{12}$ G, as seed, 
equipartition field strength in the central object will be reached (provided enough
kinetic energy is available, see below) in only $\approx 40$ ms. The equipartition field
strengths in the remnant are a few times $ 10^{17}$ G for the central object and around
$\sim 10^{15}$ G for the surrounding torus (see Fig. 8 in Rosswog et al. 2003).
To estimate the maximum obtainable magnetic field strength (averaged over the 
central object) we assume that all of the available kinetic energy can be 
transformed into magnetic field energy. Using the kinetic energy stored in the 
rotation of the central object, $E_{\rm kin}= 8\cdot10^{52}$ erg for our generic simulation,
we find $\langle B_{co} \rangle= 
(3\cdot E_{\rm kin}/R_{co}^3)^{1/2} \approx 3\cdot 10^{17}$ G, 
where $R_{co}$ is the radius of the central object
(note that if only a fraction of 0.1 of the equipartition pressure should be
reached this would still correspond to $\sim 10^{17}$ G).\\
There are various ways how this huge field strength could be used to produce a GRB.
The fields in the vortex rolls (see Fig. 8 in Rosswog and Davies 2002) will wind up the magnetic 
field fastest. Once the field reaches a strength close to the local equipartition value 
it will become buoyant, float up, break through the surface and possibly reconnect in 
an ultra-relativistic blast (Kluzniak and Ruderman 1998). The time structure imprinted on the 
sequence of such blasts would then reflect the activity of the fluid instabilities inside 
the central object. The expected lightcurve of the GRB would therefore be an erratic
sequence of sub-bursts with variations on millisecond time scales.

Simultaneously such an object can act as a scaled-up ``super-pulsar'' and drive out an 
ultra-relativistic wind. A similar configuration, a millisecond pulsar with a magnetic field
of a few times $10^{15}$ G, formed for example in an accretion-induced collapse, has been
suggested as a GRB-model by Usov (1992, 1994). The kinetic energy from the braking of the 
central object is mainly transformed into magnetic field energy that is frozen in the outflowing
plasma. At some stage the plasma becomes transparent to its own photons producing a blackbody 
component. Further out from the remnant the MHD-approximation breaks down and intense 
electromagnetic waves of the rotation frequency of the central engine are produced. These will transfer
their energy partly into accelerating outflowing particles to Lorentz-factors in excess of $10^6$
that can produce an afterglow via interaction with the external medium.
The other part goes into non-thermal synchro-Compton radiation with typical energies of $\sim 1$ MeV
(Usov 1994).

\subsubsection{Neutron Star Black Hole Mergers}

\begin{figure*}
\hspace*{-2cm}\includegraphics[height=.25\textheight]{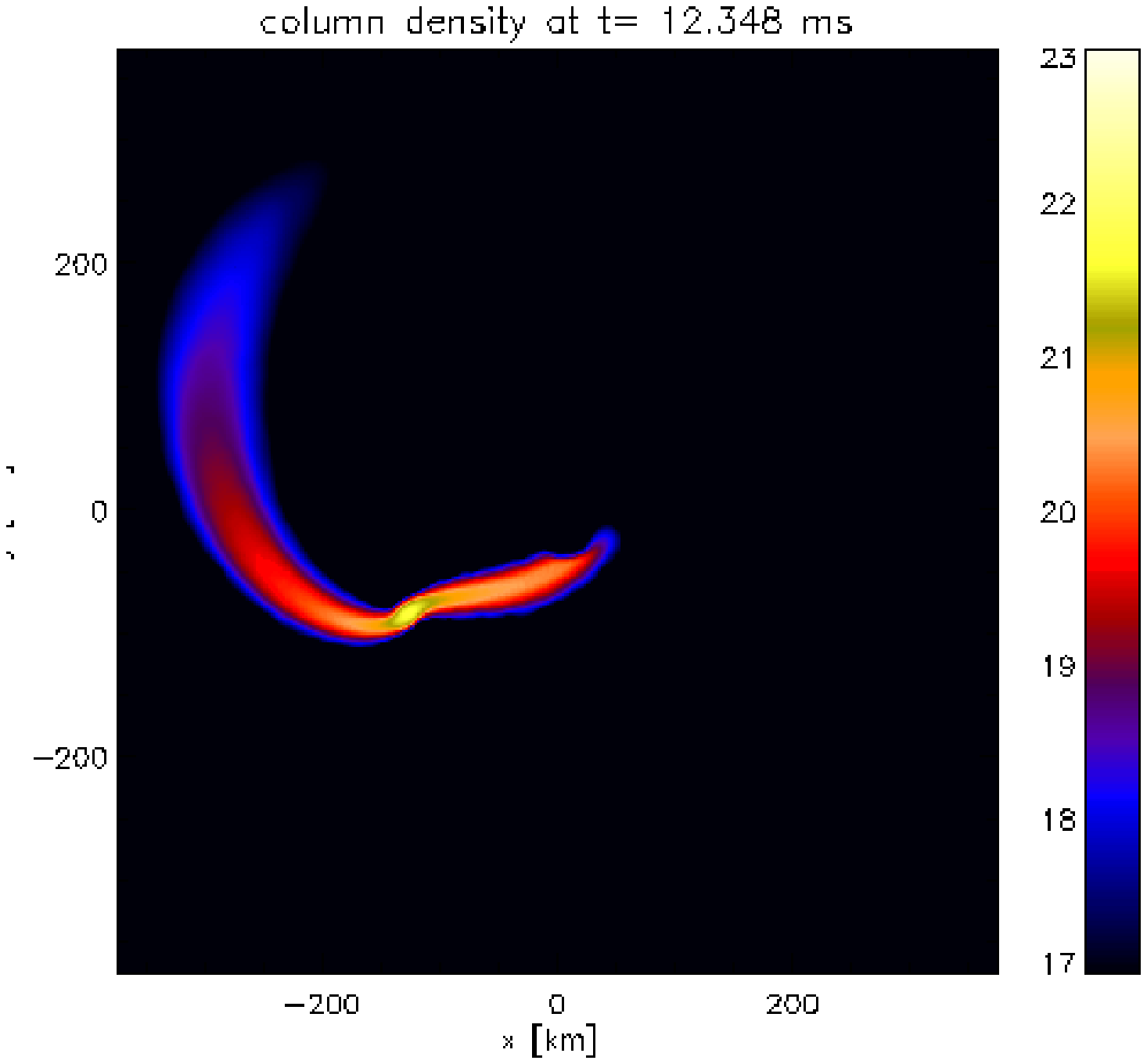}\includegraphics[height=.25\textheight]{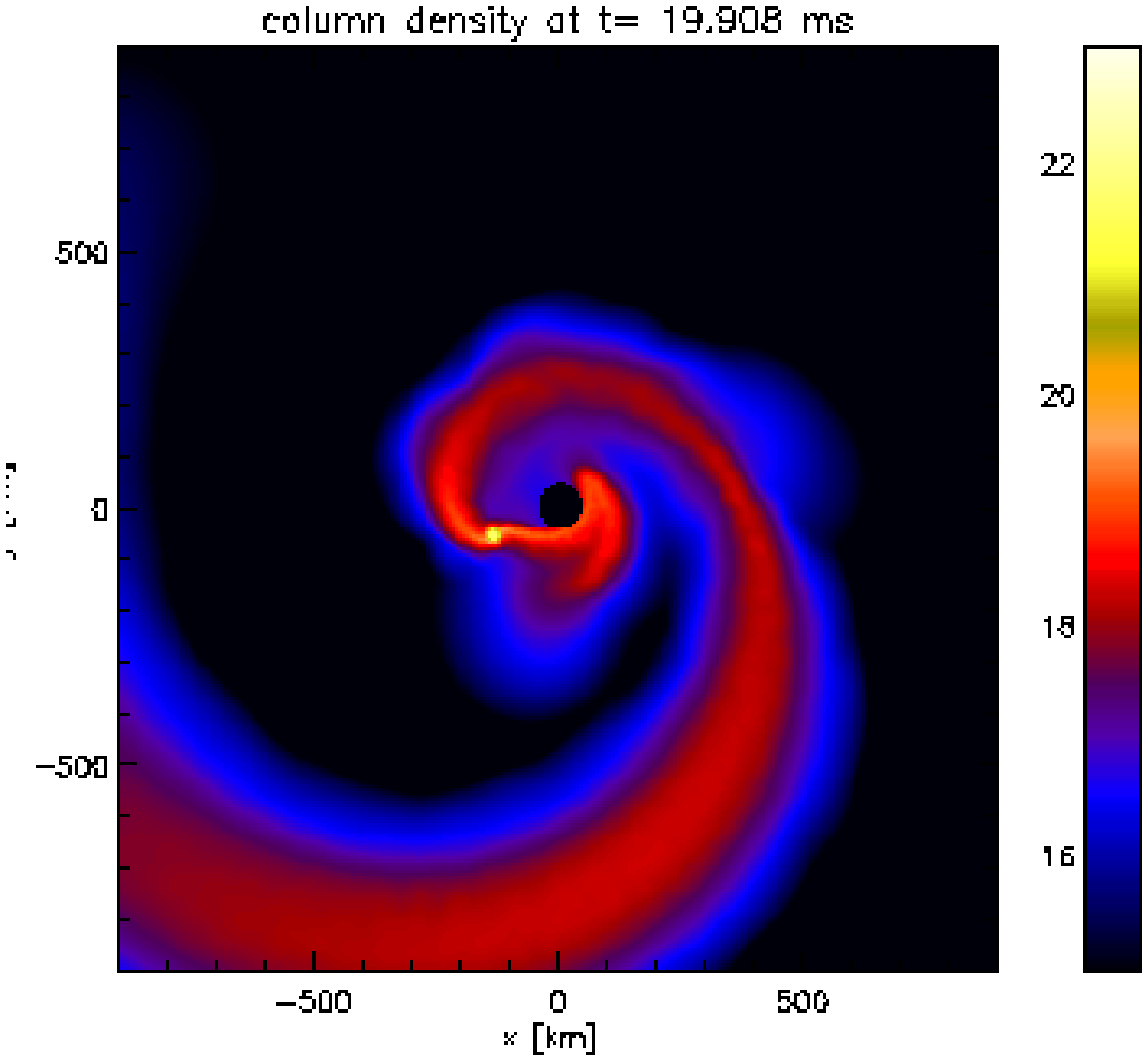}\includegraphics[height=.25\textheight]{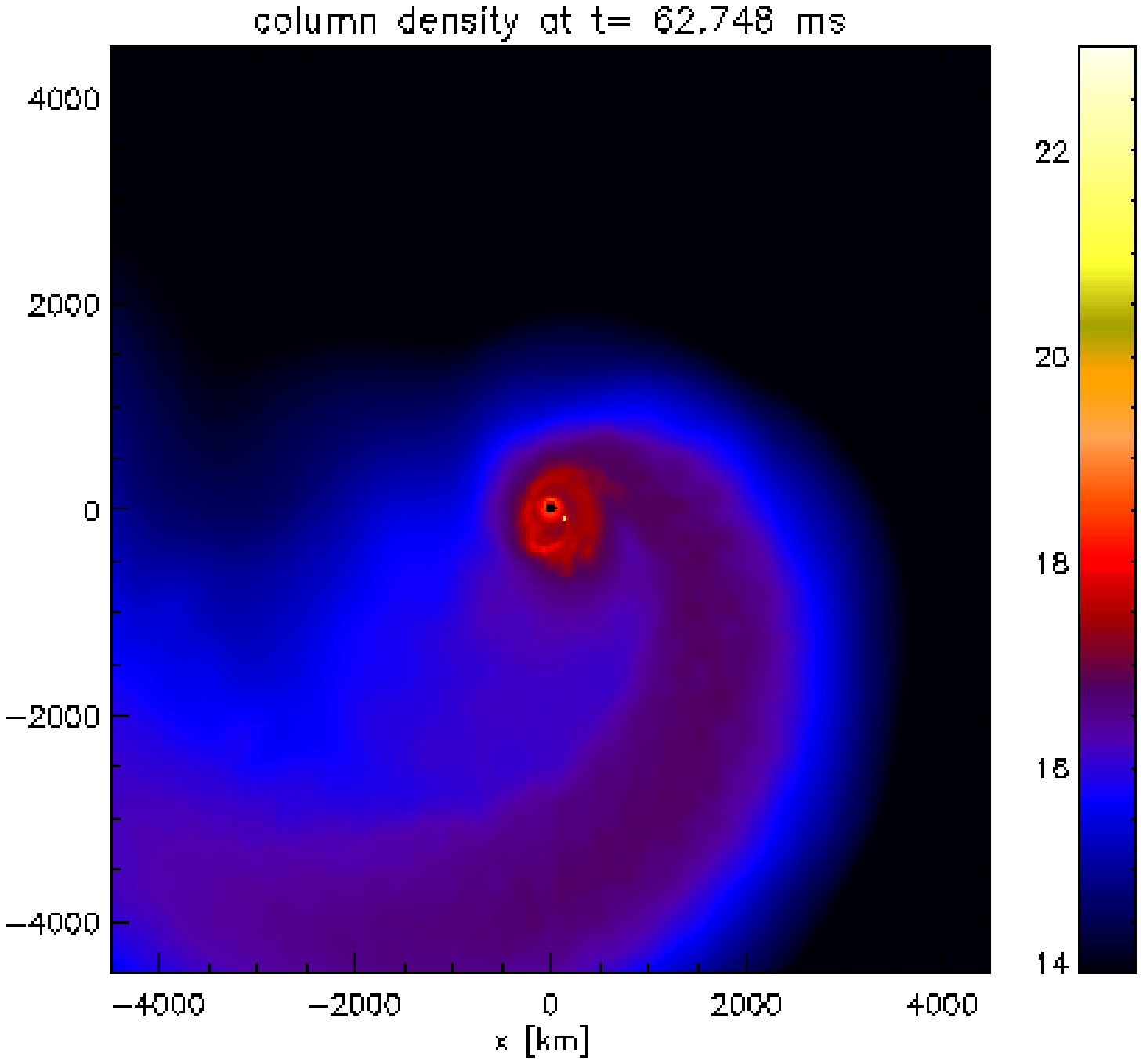}{\caption{Tidal disruption of 1.4 \msun neutron star by a 14 \msun
black hole. Colour-coded is the logarithm of the column density (g/cm$^2$). Note the change in the scales (km) of the different panels. From Rosswog et al. 2004. }
\label{NSBH}}
\end{figure*}
Neutron star black hole mergers have been estimated to occur at a rate of $10^{-4}$
(year galaxy)$^{-1}$ \cite{bethe98}, i.e. an event rate comparable to that of neutron star binaries
is expected. Moreover, their 
gravitational wave signal is stronger and therefore they could possibly dominate
the first detectable signals from inspirals of compact binaries.\\
Simulations of this scenario have been performed by Lee 
\cite{lee99a,lee99b,lee01} using Newtonian gravity and polytropic equations of state
of varying stiffness.
Ruffert, Janka and Eberl performed similar simulations but with a detailed
microphysics input (nuclear equation of state and neutrino leakage \cite{janka99}).
In our own simulations of NS-BH mergers we used a relativistic mean field equation of state
\cite{shen98} together with three-dimensional smoothed particle hydrodynamics and a detailed, multiflavour 
neutrino treatment.\\
The interaction of mass transfer, gravitational wave backreaction and the reaction of the
neutron star radius to the mass loss leads to a very complicated accretion dynamics
in a neutron star black hole system. We find in all of our simulations (apart from
an extreme test case with mass ratio $q=0.93$, i.e. a black hole of 1.5 \msun and a 
neutron star of 1.4 \msun) that a ``mini-neutron star'' survives. A specific example,
a black hole with 14 \msun, an initially tidally locked neutron star of 1.4 \msun 
modelled using Newtonian gravity plus gravitational wave backreaction forces is shown in Figure \ref{NSBH}.
We find initial peak mass transfer rates of $\sim 500$ \msun/s (at about the time shown
in the first panel of Figure \ref{NSBH}). After that phase the orbit widens, and we find 
a long sequence of mass transfer episodes, always transferring mass into an accretion disk
around the hole, moving out, spiralling in again and so on. At the end of the simulation
(after around 64 ms and seven mass transfer phases) the neutron star still has around 0.2 
\msun and shows no sign of being disrupted soon (the later mass transfer rates
reduce the neutron star mass only by small amounts). The accretion disk that forms around the hole is very low
in mass ($ > 10^{-2}$ \msun) and relatively moderate in temperature ($ T \sim$ 2 MeV).
Under these circumstances the neutrino luminosity is around an order of magnitude smaller 
than in the binary neutron star merger case and therefore not very encouraging to power
a GRB. 
Systems without initial neutron star spins yield somewhat more promising
accretion disks, but details remain to be explored further. These results are
sensitive to the stiffness of the equation of state, a softening of the 
EOS at higher densities (e.g. due to the presence of hyperons) could possibly 
substantially alter the results.

\section{Summary and Prospects}

In the last few years the GRB field has seen tremendous progress. The 
cosmological distance scale is now well-established. Moreover, since 
GRB030329 the connection of (long) GRBs with supernovae, suspected already in 
the very first paper on GRBs by Klebesadel et al. (1973), has been put to a 
firm footing. This lends further strength to the collapsar model of 
the long burst category.\\
But there are still myriards of open questions both related to the
astrophysical events/central engines as well as related to the fundamental 
physics involved. For example, is our picture of stellar evolution wrong if
state of the art progenitor models cannot produce progenitors of sufficient 
angular momentum for a collapsar? If collapsars result from Wolf-Rayet stars
then why does the expected ambient matter density distribution resulting
from a stellar wind ($\rho \propto r^{-2}$) not give better results in 
interpreting observations than a constant ambient matter density? What is 
the connection between GRBs and X-ray flashes? Is it the same event seen 
from different angles? Do short GRBs have afterglows? Why are their spectra 
harder? Do they also occur close to star froming regions?\\
Or from the physics side: by which physical mechanisms are jets launched from 
accretion disks? How are the ejecta accelerated to Lorentz factors beyond 100?
How are the magnetic fields at the emission site created? Are they advected
from the central engine or are they produced by in shocks?\\
Hopefully, at least some of the above questions can be answered in the near future by missions such as SWIFT.

\section*{Acknowledgements}
The simulations of the compact binary mergers were
performed using both the UK Astrophysical Fluids Facility (UKAFF) and
the University of Leicester Mathematical Modelling Centre's
supercomputer. S.R. gratefully acknowledges a PPARC Advanced Fellowship.

\bibliographystyle{apalike}
\chapbibliography{logic}


\end{document}